\documentclass[aps,twocolumn,floatfix]{revtex4}
\usepackage{graphicx}
\usepackage{latexsym,amsmath,amssymb}

\def\C60{C$_{60}$}

\begin{document}

\title{Glass transition line in \C60:
a mode-coupling/molecular-dynamics study}

\author{D.~Costa$^1$
\footnote{Corresponding author; email: {\tt dino.costa@unime.it}}, 
R.~Ruberto$^{1,2}$, F.~Sciortino$^3$, M.~C.~Abramo$^1$, 
and C.~Caccamo$^1$}
\affiliation{
$^1$Dipartimento di Fisica, Universit\`a di Messina and CNISM,
Ctr Papardo -- 98166 Messina, Italy \\
$^2$CNR-INFM DEMOCRITOS National Simulation Center and
Dipartimento di Fisica Teorica, Universit\`a di Trieste,
Strada Costiera 11 -- 34014 Trieste, Italy\\
$^3$Dipartimento di Fisica, Universit\`a di Roma ``La Sapienza'',
Piazzale Aldo Moro 2 -- 00185 Roma, Italy
}

\begin{abstract}
We report a study of the  mode-coupling theory (MCT) glass transition line  
for  the Girifalco model of \C60 fullerene. 
The equilibrium 
static structure factor of the model, 
the only required input for the MCT calculations, is provided by 
molecular dynamics simulations. 
The glass transition line  develops inside the metastable liquid-solid
coexistence region and extends down in temperature,
terminating on the liquid side
of the metastable portion of the liquid-vapor binodal.
The vitrification locus 
does not show re-entrant behavior.
A comparison with previous computer simulation
estimates of the location of the glass line
suggests that the theory 
accurately reproduces  the shape of
the arrest line in the density-temperature plane. 
The theoretical HNC and MHNC structure factors
(and consequently the corresponding MCT glass line)  
compare well with the numerical counterpart. These evidences
confirm the conclusion drawn in previous works
about the existence
of a glassy phase for the fullerene model at issue.
\end{abstract}

\maketitle

\section{Introduction}

The onset of a glassy phase characterized by positional disorder
in the Girifalco  central potential model of \C60~\cite{girifalco:92}
 has been recently documented
by some of us via Molecular Dynamics (MD) studies~\cite{abramo:04,abramo:05}. 
Interest in the 
vitrification process in fullerenes stems not only from the intrinsic
relevance of this class of materials, but also from the 
nature of their interparticle interaction. The Girifalco
potential  appears in fact to 
be ``marginally'' short-range, giving rise to peculiar effects 
when we consider
the interplay between the decay 
of the interactions and both the characteristics of the phase portrait 
and the glass forming ability of this model.
   In particular, following
an initial debate on the existence of a stable liquid phase for
this model~\cite{mooij:93,cheng:93}, it has been shown 
that the liquid pocket in the \C60 phase diagram is confined to a tiny
temperature interval (see e.g.~\cite{costa:03,hasegawa:99}
and references therein).  In this sense,
the system displays a characteristic
``borderline'' behavior, intermediate between 
what one expects for the phase equilibria of a simple fluid
(with a fully developed liquid phase), 
and a condition where the liquid-vapor equilibrium 
is only metastable with respect to the vapor-solid phase separation, 
the binodal curve
falling below the sublimation line.
The latter behavior
is usually observed when the range of attractive forces
is short enough compared with
the size of the repulsive
core, a condition  typically faced 
when one considers effective models for 
macrosized molecular systems, like protein 
solutions~\cite{muschol:97,rosenbaum:99,piazza:00}  or
colloidal suspensions~\cite{poon:98}.

  On the other hand, extensive studies of the glass transition in 
simple systems like square wells, adhesive hard spheres 
and hard-core Yukawa fluids~\cite{fabbian:99,dawson:00,foffi:02} 
(see ref~\cite{advances} for a recent review), 
based on the application of the Mode-Coupling Theory 
(MCT,~\cite{gotze:91}) have allowed 
to identify distinct mechanism of formation of 
glasses according to the balance
of the repulsive and attractive interactions. In particular,
systems with sufficiently short-range attraction exhibit, 
together with 
a normal repulsion-driven glass
which behaves qualitatively like a hard-sphere glass,
an ``attractive'' glass of different nature, favored both by the energy and
the local entropy~\cite{dawson:01,zaccarelli:01}.
This
circumstance naturally candidates the \C60 
model---wherein, as observed,
the subtleties related to the shape of the interaction potential
play a crucial role---for a study of the glass transition 
and of its typicality.

Besides our simulations investigations~\cite{abramo:04, abramo:05},
the study of the glass transition line in the Girifalco model
has been recently 
addressed by Greenall and Voigtmann~\cite{greenall:06}. 
In ref~\cite{greenall:06} these authors carry out 
ideal MCT calculations, using as input data for the theory 
the static structure factors $S(k)$ obtained from the
Hypernetted Chain~(HNC) and Percus-Yevick~(PY)
 liquid state theories~\cite{TOSL}.
They show that
vitrification in the \C60 model occurs,
in agreement with MD results,
although at densities lower than those
predicted by the computer simulations;
such an underestimate is not unexpected
given the inherent inaccuracies of MCT~\cite{foffi:02}.
Moreover,
the features of the MCT non-ergodicity parameter
and the overall behavior of the glass
transition line indicate the crossover to an attractive-glass 
behavior at relatively low temperatures, thereby expanding the
scenario emerging from previous simulations.
MD simulations 
have shown in fact 
evidences of a repulsive glass only,
over the whole temperature range investigated~\cite{abramo:04,abramo:05}.
The possibility that an attractive glass can exist
for the Girifalco model
appears somewhat unexpected on the basis
of the broad analysis carried out in ref~\cite{foffi:02}, 
in which a Yukawa model with parameters 
compatible with the decay rate of the Girifalco potential
displays a repulsive glass only.
A similar conclusion is drawn in ref~\cite{greenall:06} itself:
if one uses a square-well potential
mimicking the attraction range of the Girifalco model,
the resulting interaction is not short-range enough
to determine the appearance of an attraction-driven glass.

The possibility
that the attractive interaction
in the Girifalco model is sufficiently short-range 
to display several peculiar 
features of an attractive glass
poses intriguing questions about the whole mechanism
which underlies the existence of such an arrested state 
of matter, with implications for a variety of similar potentials
currently used in colloid and protein studies.
On the other hand,
the arguments put forward in ref~\cite{greenall:06} hinge 
on two approximate liquid state approaches, like the HNC and the PY theories,
and on a previous
analysis carried out by the same authors~\cite{greenall:06b}, 
about the (weak) sensitivity of MCT predictions
to the $S(k)$ behaviour in the low-$k$ region, below the first
diffraction peak.
This scenario motivates in our opinion
an investigation of MCT predictions implemented through the use 
of accurate structure factors.  We have performed to this aim
extensive MD calculations 
in the temperature and density regimes inside the vapor-solid
and liquid-solid regions, and extracted the structure factor directly
from the simulation data, down to the lowest $k$ vector compatible with
the simulation box size. 
In order to assess the theoretical predictions,
we have also calculated the static structure factor
in some selected thermodynamic state points,
in the framework of the 
the Modified HNC (MHNC,~\cite{MHNC}) approach, 
a theory which yields
accurate results for the \C60 model~\cite{caccamo:95}.

The paper is organized as follows: 
sect.~II is devoted to an introduction of the model, of the simulations
strategy and of the basic equations employed in the MCT. In sec.~III 
results are reported and discussed, whereas 
sect.~IV contains a few concluding remarks.

\section{Model, simulation procedure and MCT approach}

The Girifalco interaction 
potential $v(r) $ among two \C60 molecules reads~\cite{girifalco:92}: 
\begin{eqnarray}
\label{eq:pot}
v(r)  & = &    -\alpha_1 \left[ \frac{1}{s(s-1)^3} +
                   \frac{1}{s(s+1)^3} - \frac{2}{s^4} \right] 
\nonumber\\[4pt] & & \quad   
+ \alpha_2    \left[ \frac{1}{s(s-1)^9} +
                   \frac{1}{s(s+1)^9} - \frac{2}{s^{10}} \right]~~
\end{eqnarray}
where $s=r/d$, $\alpha_1=N^2A/12d^6$, and
$\alpha_2=N^2B/90d^{12}$;
$N=60$ and $d=0.71$\,nm are the number of carbon atoms and the diameter,
respectively, of the fullerene particles;
$A=32\times10^{-60}$\,erg\,cm$^6$ and
$B=55.77\times10^{-105}$\,erg\,cm$^{12}$ are constants entering the
12-6 potential $\phi$(r)$= -A/r^6+ B/r^{12}$ through which two carbon sites on
different spherical molecules are assumed to interact.
The distance where the potential in eq~\ref{eq:pot} crosses zero,
the position of the potential well minimum and its depth,
are $\sigma=0.959$\,nm, $r_{\rm min}= 1.005$\,nm, and
$\varepsilon = 0.444\times10^{-12}$\,erg, respectively.

\begin{figure}
\begin{center}
\includegraphics[width=8cm]{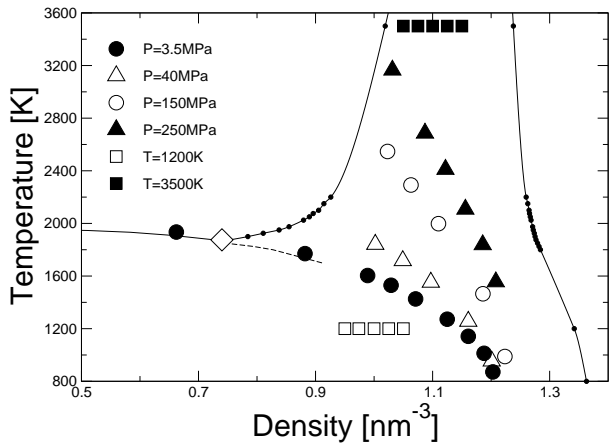}
\caption{Cooling paths at $P=3.5$, 40, 150, and 250\,MPa  and
constant temperature runs
at $T=1200$ and 3500\,K.
Full lines correspond to
the liquid branch of the binodal and the
liquid-solid coexistence lines~\cite{costa:03}. The triple point (diamond)
and the metastable portion of the binodal line (dashed curve) are
also shown. Dots collectively show all past~\cite{costa:03}
and newly added (see text) estimates of coexistence points.
}\label{fig:cooling}
\end{center}
\end{figure}

MD simulations are carried out on a system
composed of 1000 particles 
enclosed in a cubic box with periodic boundary conditions.
The Andersen algorithm~\cite{andersen:80}
 is used to enforce constant-pressure $P$, 
constant-enthalpy $H$ conditions to the sample.
We have analyzed the system along the cooling paths
characterized by the pressure 
$P=3,5$, 40, 150 and 250\,MPA, according to 
the strategy already documented in 
refs~\cite{abramo:04,abramo:05}.
As visible in Figure~\ref{fig:cooling},
five to seven 
thermodynamic states for each pressure
are selected
around the glass transition points (which have been
estimated through several structural and dynamic
indicators in ref~\cite{abramo:05}), 
and long trajectories
are generated to estimate the equilibrium static
structure factor $S(k)$ to be fed into  
the mode-coupling theory calculations. 
Several state points along the isotherms
$T=1200$\,K and $T=3500$\,K 
are also analyzed, in order
to characterize the behavior of the vitrification
curve in the limits of relatively 
low and high temperatures (see Figure~\ref{fig:cooling}).

Runs of 12\,000 time steps (with $\Delta t=5$\,fs)
are generally long enough to ensure a stable estimate of structure factors.
Two method for the calculation of $S(k)$
(namely through a direct estimate of fluctuations of 
the density $\rho$ and by fourier inversion of the radial distribution
function $g(r)$) are used and compared in this study.
Results are generally equivalent: the 
calculations coming from $g(r)$ are on the whole
less noisy than those obtained through the direct method
whereas the latter are more accurate in the small-$k$ region.
If necessary, a smoothing procedure has been applied
to the $S(k)$ data prior to mode-coupling calculations.

MCT derives equations for the normalized 
time-dependent density correlators of the Fourier 
components of the particle density 
fluctuations $\delta\rho_k(t)$~\cite{gotze:91}:
\begin{equation}
\Phi_{k}(t)
\equiv
{\frac{\left<\delta\rho_{k}(t)\delta\rho_{-{k}}(0)\right>
}{\left<{|\delta\rho_{k}|}^2\right> }}.
\end{equation}
starting only from the number density $\rho$ and the structure factor
$S_k = \left< |\delta \rho_{k}|^2 \right>/N$ 
(or equivalently the direct correlation function
$\rho c_{k} = 1 - S_{{k}}^{-1}$. The glass transition predicted 
by MCT is obtained  solving the $t \rightarrow \infty$ limit of 
the equations for the normalized correlators $\Phi_{k}(t)$, 
the so-called non-ergodicity factor $f_{k}$:
\begin{equation}
f_{k} = \lim_{t \rightarrow \infty} \Phi_{k}(t).
\end{equation}
The equations have the form:
\begin{equation}
\lim_{t \rightarrow \infty} m_{k}(t) = \frac{f_k}{1-f_k}
\label{eq:mem1}
\end{equation}
where the memory kernel is given by
\begin{eqnarray}
\label{eq:mem}
m_{k}(t) &=&  \frac{\rho}{V}
\sum_{k' \neq k} S_k S_{|{k}-{k}'|} S_{{k}'} \times \nonumber \\
& &  \quad
\left|
{\frac{{\bf k}\cdot{\bf k'}}{k^2}}\ c_{k'} +
{\frac{{\bf k}\cdot({\bf k}-{\bf k'})}{k^2}}\ c_{{k}-{k'}}
\right|^2 \times \nonumber \\
& & \qquad \Phi_{|{k}-{k}'|}(t) \Phi_{{k}'}(t)
\end{eqnarray}
For specific values of the input parameters,
the solution to these equations admits not only the usual trivial solution $f_k = 0$, but also solutions with $f_k \neq 0$. The value of $f_k$ at the transition point is denoted $f_k^c$.  
We have solved equations~\ref{eq:mem1}-\ref{eq:mem} 
on a grid of 300 wavevectors up to $k=32$\,nm$^{-1}$
using a standard iterative procedure.

\begin{figure}[!t]
\begin{center}
\begin{tabular}{c}
\includegraphics[width=7cm,angle=-90]{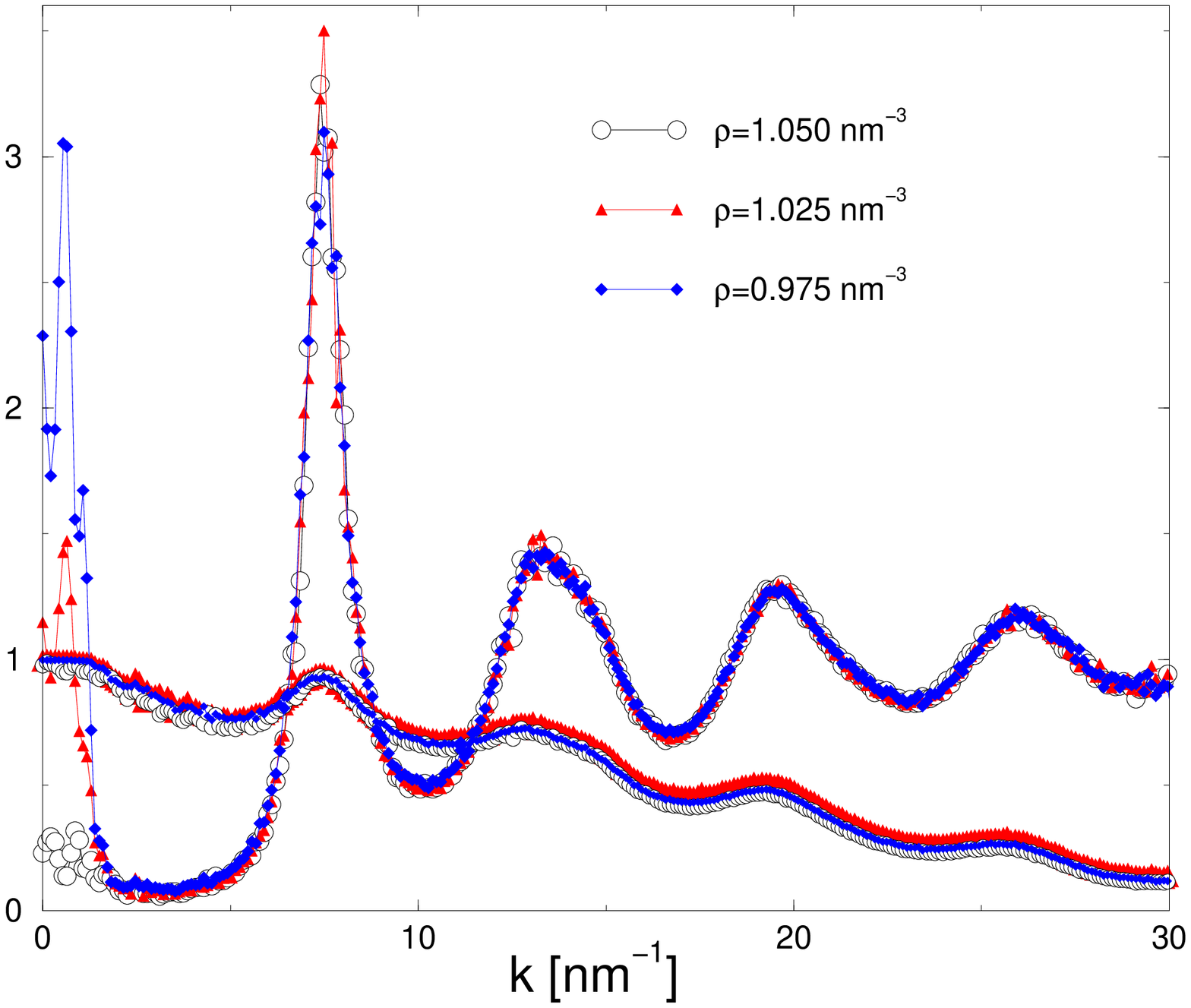} \\
\includegraphics[width=7.2cm,angle=-90]{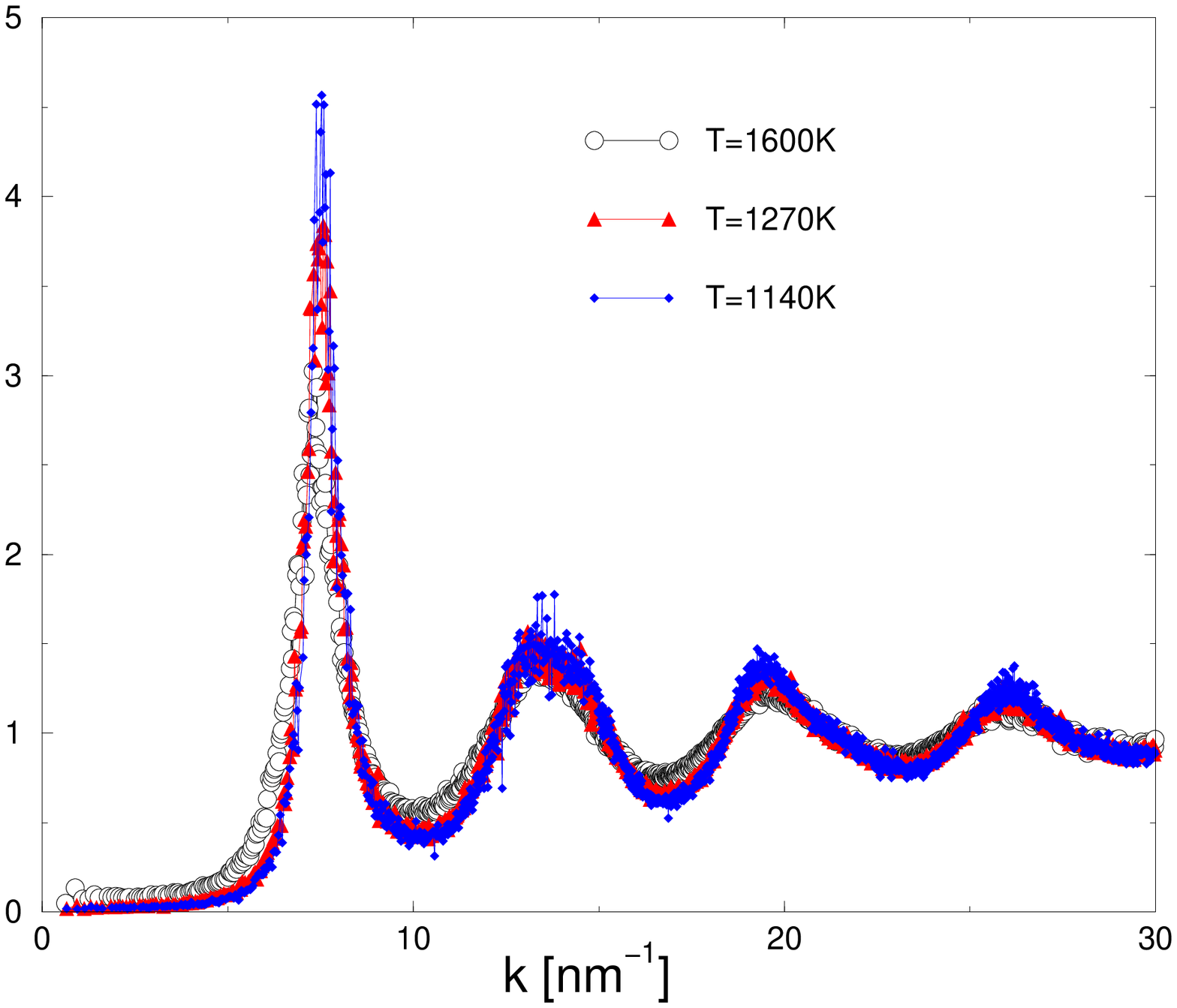}
\end{tabular}
\caption{MD static structure factors
along the $T=1200$\,K  (left)
and $P=3.5$\,MPa (right) paths.
In the top panel, the corresponding non-ergodicity factors
are also displayed.}
\label{fig:sk}
\end{center}
\end{figure}

\section{Results}

An overview of all state 
points encompassed in this work is reported in Figure~\ref{fig:cooling},
in the context of the phase diagram of the \C60 model
calculated in ref~\cite{costa:03}.
Newly generated fluid-solid
coexistence points in the high temperature regime ($T = 3500$)
and in the low temperature solid phase ($T \le 1200$)
are displayed as well, in order to elucidate the whole appearance
of the coexistence region where all calculations have been done.
The coexistence points at high temperature are in particular calculated 
according to the procedure 
employed in ref~\cite{costa:03b}: the free energy of 
the fluid phase is calculated through thermodynamic integration
of the MD pressure at several state points
along the isotherm $T=3500$\,K,
whereas for the solid phase we have used a 
first-order perturbation theory
starting from a crystal of hard spheres, whose diameter is chosen
according to the Weeks-Chandler-Andersen (WCA) procedure~\cite{weeks:71}.   
We have obtained in this way for the coexisting densities: 
$\rho_{\rm fluid}(T=3500)=1.02$\,nm$^{-3}$ 
and $\rho_{\rm solid}(T=3500)=1.24$\,nm$^{-3}$. 
As for the isotherms $T=1200$\,K and $T=800$\,K, 
we have assumed that the coexisting solid density 
must be that where the (perturbation theory) pressure
reduces to zero---a simple approximation justified by the fact
that in this regime the coexisting vapor density is extremely low 
and thus it plausibly corresponds to almost zero pressure.
We have obtained:
$\rho_{\rm solid}(T=1200)=1.34$\,nm$^{-3}$ and 
$\rho_{\rm solid}(T=800)=1.36$\,nm$^{-3}$.

\begin{figure}
\begin{center}
\begin{tabular}{c}
\includegraphics[width=8.0cm]{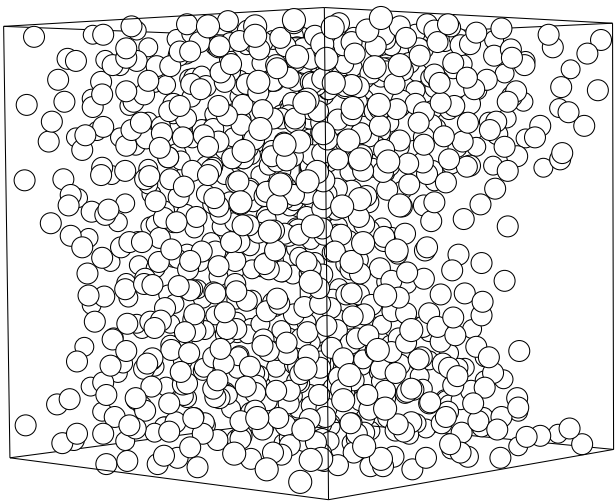}
\end{tabular}
\caption{Snapshot of the final configuration of the system
at $T=1200$\,K and $\rho=0.95$\,nm$^{-3}$; a void region
clearly displays in the sample. }
\label{fig:conf}
\end{center}
\end{figure}

\begin{figure}
\begin{center}
\begin{tabular}{c}
\\[10pt]
\includegraphics[width=7.5cm]{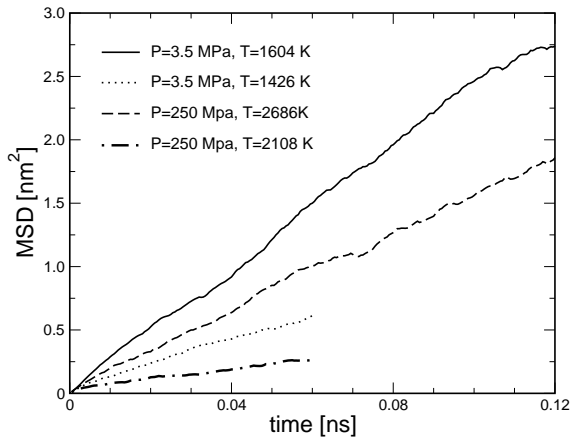}
\end{tabular}
\caption{Mean square displacements across the MCT vitrification
thresholds at $P=3.5$ and 250\,MPa. Longer runs
correspond to the states immediately before the transition point.
}\label{fig:msd}
\end{center}
\end{figure}

As visible from Figure~\ref{fig:cooling}, 
thermodynamic states at pressures increasingly higher 
than $P=3.5$\,MPa are allocated between the 
freezing and the melting lines of the model.
State points along the isobaric path $P=3.5$\,MPa
are almost superimposed on the liquid branch of the metastable portion
of the liquid-vapor phase separation, whereas all
thermodynamic states along the isotherm $T=1200$\,K
are definitely inside the vapor-solid coexistence region. 
This point can be further illustrated by the evaluation
of the structure factors at $T=1200$\,K and $P=3.5$\,MPa,
reported in Figure~\ref{fig:sk}. 
It appears that the structure factor at $T=1200$\,K
is barely sensitive to density variations in
the range 0.95---1.05\,nm$^{-3}$, but
for the steep rise of the $k\to0$ limit when the density is decreased,
clearly indicating a more and more pronounced tendency of the 
sample to phase separate as it goes deeper and deeper inside
the (metastable) binodal. 
On the other hand, the structure factor
at $P=3.5$\,MPa keeps around zero in the $k\to 0$ limit
and becomes more structured during the cooling sequence
documenting  in this way the approach to the solid configuration.

In parallel, 
the spatial configurations obtained at $T=1200$\,K,
analyzed in terms of nearest-neighbors distributions of particles
and by means of a direct visual inspection,
reveals at $\rho=0.95$\,nm$^{-3}$
the existence of a rarefact region in the sample 
surrounded by a uniformly denser environment (see the snapshot 
in Figure~\ref{fig:conf}). Such strong
dishomogeneities  tend to persist
almost up to $\rho=1.025$\,nm$^{-3}$ and disappear  
at higher densities: at $\rho=1.05$\,nm$^{-3}$ 
we have detected 
no evidence of (incipient) phase separation,
and the system displays a uniform distribution of particles
inside the simulation box.

At variance with the latter 
regime, all other points
investigated in this work represent truly 
equilibrated thermodynamic states.
In order to be sure that all samples are properly equilibrated,
and that each particle diffuses on average more than its diameter
over the whole simulation time, 
we have recorded the mean square displacement~(MSD)
over particularly long  simulation runs for some selected state points
where MCT predicts 
the  glass transition.  Results are reported
in Figure~\ref{fig:msd}: as visible, 
observation windows of the order of $\sim 0.1$\,ns allow
the particles to diffuse for a distance varying
between $\sim 1.4$ and $\sim 1.7$ molecular
diameters, the latter being roughly identified with
the ``collisional'' distance $\sigma$ of the \C60 potential
(see section~II).  The diffusion coefficient varies between 
$\sim 2.6 \times 10^{-5}$ and $3.8 \times 10^{-5}$\,cm$^2$/s  
before the MCT transition and
drops to $\sim 1.6 \times 10^{-5}$~---~$6.6 \times 10^{-6}$\,cm$^2$/s 
thereafter.
Data in Figure~\ref{fig:msd} confirm the known MCT 
tendency to predict a glass transition at state points where
the dynamics of the real system is still ergodic.

\begin{figure}
\begin{center}
\includegraphics[width=8.0cm]{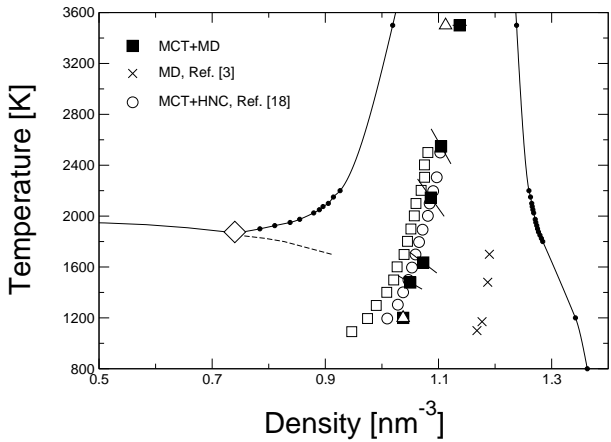}
\caption{Glass transition points as obtained in this work through
MCT calculations based on MD (full squares),
and MHNC (triangles)
structure factors.
Crosses are the glass transition
estimates through direct MD calculations~\cite{abramo:04,abramo:05}.
MCT predictions with HNC (circles)
and PY (open squares)
structure factors~\cite{greenall:06}
are also shown.
}\label{fig:glass}
\end{center}
\end{figure}

\begin{figure}[!ht]
\begin{center}
\begin{tabular}{c}
\includegraphics[width=6.5cm,angle=-90]{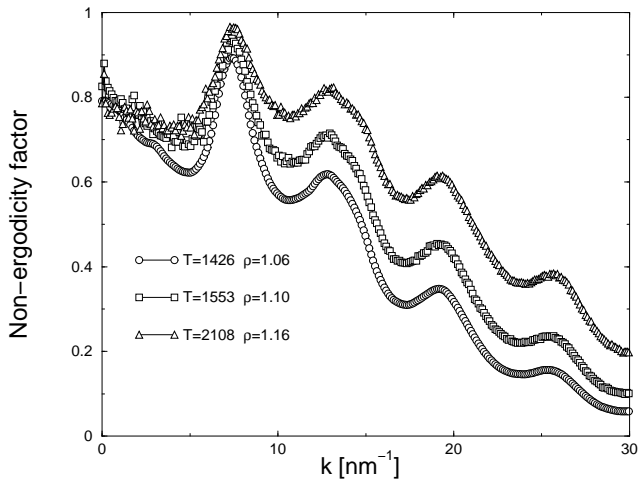}
\end{tabular}
\caption{Non-zero ergodicity factors at the MCT
transition thresholds ($f^c_k$)
for  pressures $P=250$, 40 and  3.5\,MPa (from top to bottom).
State points along such isobaric paths
with temperatures immediately higher
than those reported in the figure
are still ergodic, i.e. $f_k=0$.
Densities and temperatures in the legends
are expressed in nm$^{-3}$ and K, respectively.
}\label{fig:nefc}
\end{center}
\end{figure}

Main results of this work are presented in 
Figure~\ref{fig:glass}, where the glass transition line identified through 
MCT calculations is displayed. We also report in the same figure
the theoretical predictions from MHNC calculations
at $T=1200$\,K and 3500\,K, 
our previous MD data for vitrification~\cite{abramo:04,abramo:05}, and
the estimate of ref~\cite{greenall:06}.
MHNC calculations follow 
the general approach already employed successfully
to predict the liquid-vapor coexistence of 
the \C60 model at issue~\cite{caccamo:95}.
The numerical solutions of the MCT long-time limit equations 
(the non-ergodicity factors) across the vitrification 
thresholds are shown in Figure~\ref{fig:nefc}, 
for several isobaric paths investigated in this work.  
The last ergodic state points ($f_k=0$) and the first
non-ergodic ones ($f_k \ne 0$) are assumed to 
bracket the ideal MCT 
glass transition line reported in Figure~\ref{fig:glass}.  
The distances in the $\rho$-$T$ plane
among these couples of points hence constitute the error bars
of our predictions.

It appears from Figure~\ref{fig:glass} that the 
MCT glass transition line
does not show re-entrant behavior and 
is reasonably parallel to
the vitrification locus obtained in our previous MD simulation
study~\cite{abramo:04,abramo:05} (we shall further comment
about these specific features of our results), and runs fairly close
to the one determined through MCT
and HNC or PY structure factors as input data in ref~\cite{greenall:06}.
In particular, PY predictions tend to slightly
underestimate the vitrification density and the  
HNC results are generally
more accurate over the temperature regime investigated.

\begin{figure}
\begin{center}
\begin{tabular}{c}
\includegraphics[angle=-90,width=8.0cm]{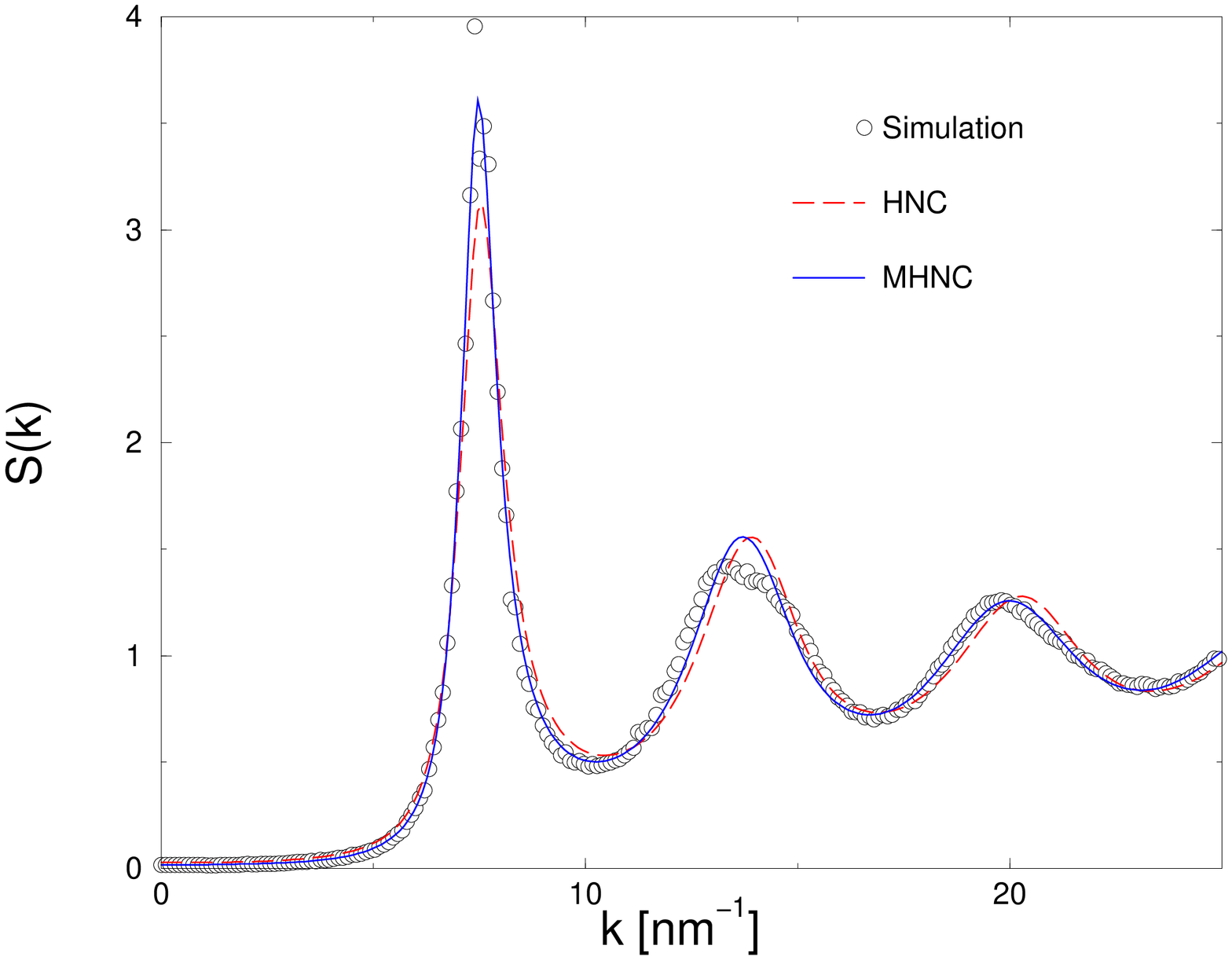} \\
\includegraphics[angle=-90,width=8.0cm]{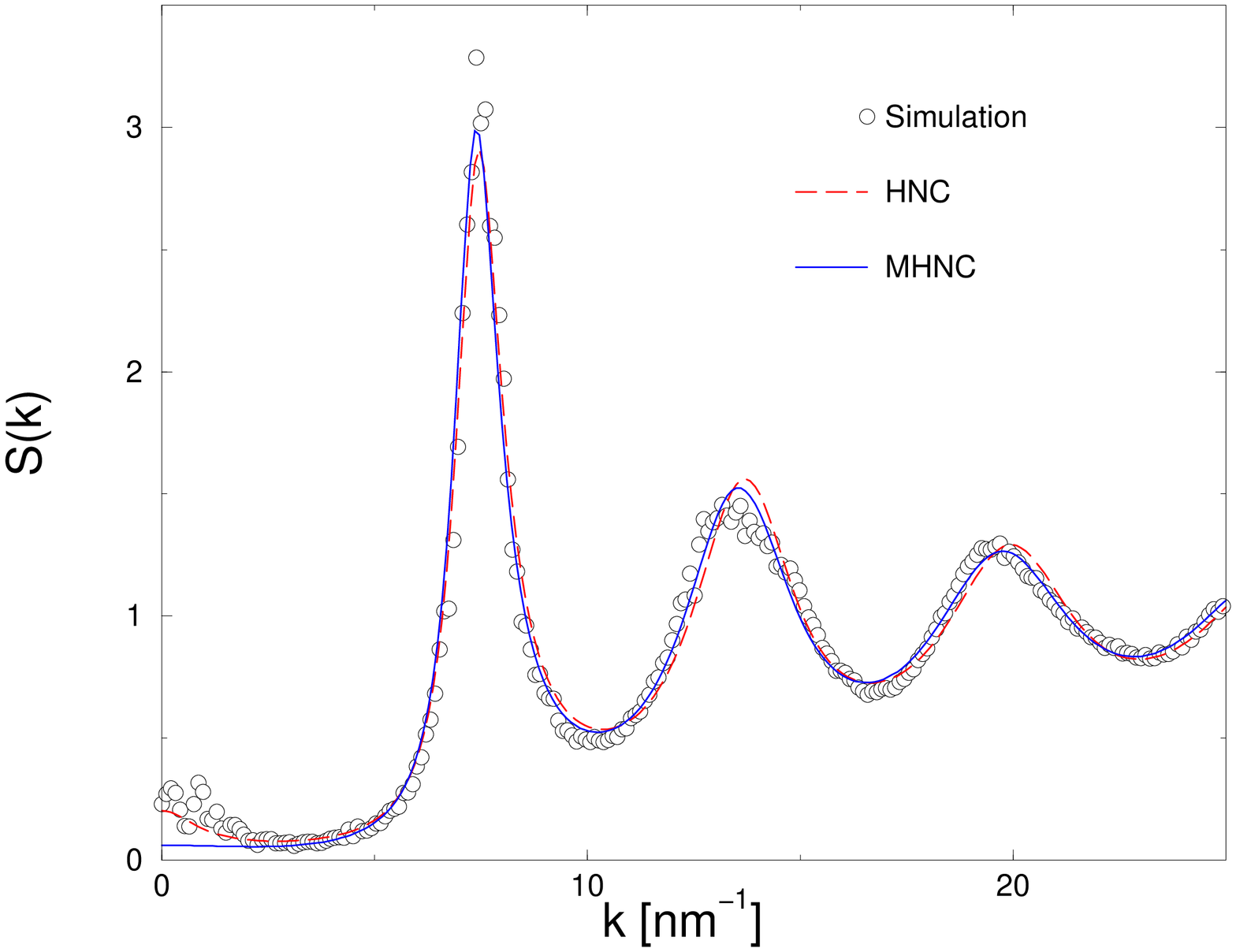} \\
\end{tabular}
\caption{Comparison between
theoretical and simulation results for
the static structure factors
at $T=3500$\,K and $\rho=1.15$\,nm$^{-3}$ (top) and
$T=1200$\,K and $\rho=1.050$\,nm$^{-3}$ (bottom).
}\label{fig:sk_3500_1150}
\end{center}
\end{figure}

Our test MHNC and HNC calculations, limited to some selected densities 
along the
isotherms $T=3500$\,K and $T=1200$\,K are also reported
in Figure~\ref{fig:glass} and  confirm the trend illustrated above. 
A comparison between 
theoretical and simulation structure factors 
is reported  in Figure~\ref{fig:sk_3500_1150}.
We display in particular data at the highest temperature and 
density investigated in this work, namely
$T=3500$\,K and $\rho=1.15$\,nm$^{-3}$, 
and data at the opposite extremum, i.e. $T=1200$\,K and 
$\rho=1.05$\,nm$^{-3}$ (the 
lowest state point which
 we have been able to investigate before the system 
displays a pronounced tendency to phase separate).
Apart from the well-known difficulties of the theoretical tools employed
to closely follow  the $k \to 0$ behavior of the structure factor, especially
in the proximity of the phase separation, it appears 
that 
both theories slightly underestimate the height
of the main peak (with MHNC theory performing  
better than HNC), and a small dephasing  
emerges in the position of the secondary peaks at higher $k$ vectors.
On the whole, we judge  
the agreement between theory and simulations 
quite satisfactory,
especially if we take into account that 
all data recorded in Figure~\ref{fig:sk_3500_1150}
have been obtained by pushing the integral equation scheme toward 
``extreme conditions'', as far as the regime in which
such tools are known to give the best performances is concerned.

Figures~\ref{fig:glass} and~\ref{fig:sk_3500_1150}
document
the good agreement between present simulation results,
our thoretical predictions, and those obtained by 
Greenall and Voigtmann~\cite{greenall:06},
equally from the point 
of view of MCT transition threshold,
and at the level of input structure factor calculations.
Our evidences indirectly  
support the conclusions drawn in ref~\cite{greenall:06}
that, 
in the framework provided by the mode-coupling theory,
and in the thermodynamic region $T \lesssim 3000$\,K,
the attractive part
of the Girifalco potential is short-range enough to 
begin to influence 
the approach to the structural arrest of the system.
The arguments presented in ref~\cite{greenall:06}
are based in particular to an extended analysis of the properties
of the non-ergodicity factor, together with a comparison
between the MCT lines obtained with either the full
\C60 potential, or with a truncated version, 
where the attractive part is cut off according to the WCA procedure.

We now turn to the examination of the low-temperature regime
MCT results.
As visible in the top panel of Figure~\ref{fig:sk},
at $T=1200$\,K  MCT predicts
the existence of a glassy phase for $\rho=1.05$\,nm$^{-3}$,
which corresponds, as discussed above,
to the lowest
density for which a genuine homogeneous
sample is observed during the time of our MD simulations, whereas
the tendency of the system to phase
separate is already well under way
at $\rho=1.025$\,nm$^{-3}$.
For this reason, and 
in view of the fact that MCT itself strictly deals with
homogeneous systems, we estimate the 
vitrification density at $T=1200$\,K 
as falling halfway between $\rho=1.025$\,nm$^{-3}$
and $\rho=1.05$\,nm$^{-3}$; this value, reported in Figure~\ref{fig:glass} 
nicely coincides with our MHNC estimates.
If lower density---but phase-separated---MD structure factors
are fed into the MCT equations, 
glass states are equally
predicted. The reason of such an outcome can be appreciated 
if we observe, again in Figure~\ref{fig:sk}, the way 
the differences in the shape of the various structure factors
reflect in 
the corresponding non-ergodicity factors:
it appears that, 
in passing from $\rho=1.050$\,nm$^{-3}$
to $\rho=0.950$\,nm$^{-3}$, the structure factor
displays minor differences, 
(especially in the 
sensitive region of the nearest-neighbor peak), 
except for
the already commented  steep rise at $k=0$.
By contrast,
the non-ergodicity factors, also shown in  
the same figure, are hardly affected by the density variation,
and infact
run almost superimposed on top of each other, thus
predicting a 
glassy phase all along the isotherm $T=1200$\,K.
 These evidences support the 
 theoretical observation that MCT predictions
 for glasses characterized by
 short-range attractions,
 are barely sensitive to the 
 low-$k$ behavior of the structure factor~\cite{greenall:06b}.

The non-zero ergodicity factors displayed in Figure~\ref{fig:sk}
would allow us to shift to lower densities the $T=1200$\,K
MCT vitrification threshold, causing a net
bending of the glass transition line, even more
pronounced than that predicted in ref~\cite{greenall:06}. 
Although
this feature would indicate an enhancement of 
the attraction-driven properties of the glass phase,
the tendency to phase separate 
of our sample detected during MD simulations
prompts us to some caution as far as 
any definite conclusion about this point is concerned. 
Similar problems
with the phase separation of the sample prevented us
from investigating the vitrification
threshold for temperatures $T<1200$\,K.
More extensive simulations,
accompanied by an analysis of the size dependence of the results,
might be worth in this respect. 
Such a  program however is unlikely to
be implemented at ease, given the fact that the thermodynamic
region to be explored falls well beneath the metastable liquid-vapor
binodal line of \C60, and definitely inside
the solid-vapor metastable equilibrium region, both circumstances
implying a strong trend to phase separation.

\begin{table}
\caption{Comparison, for several temperatures,
among the MCT transition densities
calculated in this work (column 2), the MD results of 
refs~\cite{abramo:04, abramo:05} (column 3) and the estimate which are
obtained if we apply a 12\% shift to MCT data (colunm 4).
Temperatures are given in K, densities in nm$^{-3}$.}
\label{tab:1}
\begin{center}
\begin{tabular*}{0.40\textwidth}{@{\extracolsep{\fill}}cccc}
\hline 
$T$ & MCT    &   MD  &   MCT (shifted)\\
1635 & 1.073 & 1.190 & 1.206 \\
1478 & 1.050 & 1.187 & 1.180 \\
1200 & 1.037 & 1.177 & 1.166  \\
\hline
\end{tabular*}
\end{center}
\end{table}

In comparison with our previous MD 
estimates of the glass transition of the 
Girifalco \C60 model~\cite{abramo:04,abramo:05}, 
based on several
structural and dynamic indicators,
the MCT predictions moderately underestimate
the transition densities.
It is well known that a corresponding underestimate
occurs when reference 
MCT predictions for hard spheres 
($\eta_{\rm glass}=\pi/6 \rho\sigma^3=0.516$,~\cite{franosch:97}) 
are compared with the rigorous computer simulation
result ($\eta_{\rm glass}=0.58$,~\cite{woodcock:81}).
In this case, a common procedure has been to exploit
such a $\sim 12\%$ discrepancy, 
in order to sistematically shift to higher densities
the MCT vitrification curves, also for other models,
when a comparison with simulation data is carried out
(see e.g. ref~\cite{foffi:02}). 
If we apply the same correction to our MCT points,
we obtain the remarkable agreement
with MD estimates reported in Table~\ref{tab:1}.
In refs~\cite{abramo:04,abramo:05} we have  
shown that all vitrification 
densities reported in Figure~\ref{fig:glass}
nicely fall on a single value
($(\pi/6 \,\rho \sigma_{\rm eff}^3 \simeq 0.58$, 
almost coincident with the hard-sphere
glass transition density), on condition that
all data are rescaled through the effective (temperature dependent) 
diameter $\sigma_{\rm eff}$ which is obtained by substituting the 
soft repulsive part of the Girifalco potential
with a pure hard-core  exclusion, according to the 
WCA prescription.  
As a consequence, we have deduced in refs~\cite{abramo:04,abramo:05}
that the hard-sphere behavior plays the main role in 
determining the structure of the \C60 glass, which hence results
essentially repulsion-driven in nature.
In order to reconcile the attractive nature of 
the MCT glass discussed in ref~\cite{greenall:06}
with the results of refs~\cite{abramo:04,abramo:05}
it is worth observing in Figure~\ref{fig:glass}
that in the low temperatures regime
the bending of the MD curve
is milder, with respect to the MCT predictions.
In this case, it could be possible that, 
on lowering the temperature further,
also the glass observed in our previous studies
would display a more pronounced attraction-driven effects.
On the other hand, the analysis carried out in this work
indicates that  
the strong tendency of the system to phase separate
would likely preclude the possibility of extracting meaningful 
and robust conclusion from an investigation
of the behaviour of the \C60 model in this (highly metastable)
low-temperature, high-density regime.

\section{Conclusions}

We have reported MCT determination of the glass transition line 
in a model \C60. The theoretical calculations are based
on the use, as input data for the MCT equations, of structure factors 
obtained via extensive molecular dynamics simulation investigation 
of the fullerene systems at various pressures.
    The shape of the vitrification locus is
in fairly good agreement with our previous determinations
of the glass transition entirely based 
on MD simulations~\cite{abramo:04,abramo:05}. 
In fact, the MCT predictions appear to overall underestimate the glass density
by $\sim 10$\% with respect to refs~\cite{abramo:04,abramo:05}, 
manifesting an otherwise well-known inaccuracy of the theory. 
   Present results are also in fairly good
agreement with the glass transition line 
obtained by other authors
through MCT calculations 
based on HNC structure factors input~\cite{greenall:06}.  
MCT calculations based on the 
refined modified HNC theory $S(k)$, also presented in this work,
also reproduce faithfully the glass transition line.  

As far as the interplay between 
the non-ergodicity factors and the structure
factors $S(k)$ is concerned, our evidence lends support to the 
theoretical analysis of ref~\cite{greenall:06}, where 
the 
influence of the attractive
part of the \C60 potential  on  
the structural arrest properties of the model is argued.
Our simulations document a strong tendency of the system
to phase separate 
in the low-temperature region 
 where the MCT glass transition line should exhibit
 a more pronounced attraction-driven character.
In this case, as required by MCT,  we have restricted our analysis
only to those temperatures and densities
for which a fully homogeneous sample is obtained,  
this choice resulting in a glass line with rather a moderate 
bending. More extensive computer simulation investigation,
which take into account for instance size effects, 
 might be worth in order to further enlight this point. 

The picture emerging
from the overall shape 
and characters of the glass transition line vs the 
repulsive-attractive potential features
turn out to be consistent with
the more general proposition that the glass line 
terminates on the liquid side of the
liquid-gas coexistence, when particles interact with spherical 
potentials in which the excluded volume
repulsion is complemented by attraction~\cite{sastry,foffi}. 
This scenario actually appears to be valid independently from the
range of the attractive potential~\cite{foffi}. 
Non-spherical patchy potentials, in which the number of interacting particles
is significantly reduced as compared to the spherical case,  
are necessary in order to 
suppress and shift to small densities  
the liquid-gas phase separation curve~\cite{zacca,bianchi} and hence
to extend the dynamic arrest line to lower 
temperatures and smaller densities.

\section*{acknowledgments}
This work has been done in the framework of the Marie Curie Network
on Dynamical Arrest of Soft Matter and Colloids,
Contract Nr MRTN-CT-2003-504712.

\end{document}